\documentclass[%
reprint,
prb,
superscriptaddress,
 amsmath,amssymb,
 aps,
]{revtex4-2}

\usepackage{graphicx}
\usepackage{dcolumn}
\usepackage{bm}
\usepackage{amsmath}
\usepackage{amsfonts}
\usepackage{amssymb}

\usepackage[colorlinks=true,linkcolor=blue]{hyperref}

\usepackage[utf8]{inputenc}
\usepackage[T1]{fontenc}
\usepackage{mathptmx}
\usepackage{etoolbox}

\begin{document}

\preprint{APS/123-QED}

\title{Magnetic Order and Long-Range Interactions in Mesoscopic Ising Chains}

\author{Christina Vantaraki}
\email[Corresponding author: ]{christina.vantaraki@physics.uu.se}
\affiliation{Department of Physics and Astronomy, Uppsala University, Box 516, 75120 Uppsala, Sweden}

\author{Matías P. Grassi}
\affiliation{Department of Physics and Astronomy, Uppsala University, Box 516, 75120 Uppsala, Sweden}

\author{Kristina Ignatova}
\affiliation{Science Institute, University of Iceland, Dunhaga 3, Reykjavik 107, Iceland}

\author{Michael Foerster}
\affiliation{ALBA Synchrotron Light Facility, 08290-Cerdanyola del Valles, Barcelona, Spain}

\author{Unnar B. Arnalds}
\affiliation{Science Institute, University of Iceland, Dunhaga 3, Reykjavik 107, Iceland}

\author{Daniel Primetzhofer}
\affiliation{Department of Physics and Astronomy, Uppsala University, Box 516, 75120 Uppsala, Sweden}

\author{Vassilios Kapaklis}
\email{vassilios.kapaklis@physics.uu.se}
\affiliation{Department of Physics and Astronomy, Uppsala University, Box 516, 75120 Uppsala, Sweden}

\begin{abstract}


We investigate the design of magnetic ordering in one-dimensional mesoscopic magnetic Ising chains by modulating long-range interactions. These interactions are affected by geometrical modifications to the chain, which adjust the energy hierarchy and the resulting magnetic ground states. Consequently, the magnetic ordering can be tuned between antiferromagnetic and dimer antiferromagnetic phases. These phases are experimentally observed in chains fabricated using both conventional electron-beam lithography and ion implantation techniques, demonstrating the feasibility of controlling magnetic properties at the mesoscale. The ability of attaining these magnetic structures by thermal annealing, underlines the potential of using such systems instead of simulated annealers in tackling combinatorial optimization tasks.

\end{abstract}

\maketitle


The origin of magnetism can be traced to quantum mechanical phenomena at the atomic level. Despite this fact, it is still useful and instructive to adopt classical models for the study of magnetic properties, such as emergent magnetic ordering and dynamics. Such models build upon atomistic models, making use of mostly nearest-neighbor spin-spin interactions. Even though this approach has been extremely successful, providing deep insight into many magnetic materials, it still fails to capture subtle effects arising at boundaries and cases where beyond nearest-neighbor spin-spin interactions can not be neglected anymore \cite{Zhang:2001db}. The truncation of interactions has an impact on the received order and its finite-size scaling. Addressing this theoretically or computationally is demanding, but can successfully capture finite size effects, for example, in thin films \cite{Taroni_long_range_2010}. Similarly, beyond nearest-neighbor spin-spin interactions can strongly affect the system's ordering. In the last years, the impact of the magnetic interactions can be investigated experimentally using the artificial analogs of spin systems, the mesoscopic spin systems \cite{HeydermanLJ2013Afsn, Rougemaille:2019ef, Nisoli:2017hg}.

Mesoscopic spin systems consist of arrays of lithographically patterned magnetic elements -- \textit{mesospins} -- that interact through dipolar interactions. The shape of mesospins dictate their spin dimensionality \cite{SkovdalBjornErik2021Tcos}, while the gap between them determine the strength of the interactions. Previous studies have mainly focused on close-packed arrays of mesospins, where the magnetostatic interactions are strong, leading to the emergence of collective magnetic order \cite{CowburnR.P.2000RTMQ, ImreA2006MLGf, DigernesEinar2020Diol} and dynamic behavior \cite{SkovdalBjornErik2021Tcos,ArnaldsUnnarB2016Anlo, ArnaldsUnnarB.2014Ttin, Kapaklis:2014ea, Pohlit:2020hv}. 
Extensive efforts have been made to tailor the ordering and dynamics of these systems anew, by controlling the interactions between mesospins \cite{Rougemaille:2019ef, MacauleyGavinM.2020Tmow,NguyenV.-D.2017Ciia, LouHaifeng2023Ciip}. 
These interactions are of long-range nature whose effects have been reported for systems with in-plane \cite{NguyenV.-D.2017Ciia,PerrinYann2016EdCp, OstmanErik2018Imia, Farhan:2019kd} and out-of-plane  \cite{ZhangSheng2012Pmag,Chioar_Kagome_2016} moments.
While many studies neglect this long-range nature, the importance of considering it can change dramatically the result. For example, in \citet{Chioar_Kagome_2016} it resulted in an enrichment of the phase diagram for the classical dipolar kagome Ising antiferromagnet. Another example can be found in square artificial spin ices, where the fine-tuning of beyond nearest-neighbour interactions has been the key ingredient for the recovery of degeneracy and observation of Coulomb phases \cite{PerrinYann2016EdCp, OstmanErik2018Imia, Farhan:2019kd}.

In this work, we take advantage of the long-range nature of dipolar interactions and focus on designing the order of a mesoscopic spin system by controlling them. The specific system of choice is a quasi-infinite chain of stadium-shaped mesosopins placed side-by-side, being simple enough to identify and discern the ramifications of long-range interactions. The interactions are controlled through a geometrical modification of the lattice (Fig. \ref{Figure1} (a)). Subsequent thermal demagnetization and magnetic imaging of the chains reveals the impact of long-range interactions on the received magnetic order. 


Two samples were patterned into mesoscopic chains: 
One sample was fabricated by post-patterning performed on a film, using electron-beam lithography. This film was deposited using DC magnetron sputtering on a SiO$_{2}$ substrate and consists of a 20 nm permalloy (Py) layer with a 1.5 nm Ta adhesion layer and 3 nm Ta capping layer. This procedure results in athermal (at room temperature) mesoscopic elements that interact through air. The second sample was fabricated by implanting $^{56}$Fe$^{+}$ into a Pd film through a Cr patterned mask. The Pd film was deposited using DC magnetron sputtering on a MgO substrate and composed of a 60 nm Pd layer with 5 nm V adhesion layer. This additive fabrication yields flat Fe$_{x}$Pd$_{100-x}$ (where $x$ stands for at.\%) ferromagnetic structures interacting through the Pd matrix. A detailed description about the fabrication process can be found in \citet{vantaraki2024magneticmetamaterialsionimplantation}.

In both samples, the building blocks have a length of $L$~=~1500 nm and width $W$ = 150 nm, while the gap between two adjacent elements is $g$ = 75 nm (see Fig. \ref{Figure1}). The aspect ratio of $L/W$ = 10 ensures a strong shape anisotropy, thereby forcing the magnetization of the elements to lie along their long axis, whereas the short distance between the building blocks enables sizable interactions between the elements. A geometrical vertical shift $s$ is applied to every second mesospin as a way of tuning the coupling strengths between the elements. Six different chains with $s/L$ = 0\%, 8.7\%, 25\%, 50\%, 75\%, and 100\% are studied. Thirty-three chains of the same $s$ were fabricated with a distance of 1.5~$\mu$m between them in the vertical direction to prevent interchain coupling, while each chain contains approximately 450 mesospins, mimicking a quasi-infinite chain. 

\begin{figure}[t]
\includegraphics[width=1\linewidth]{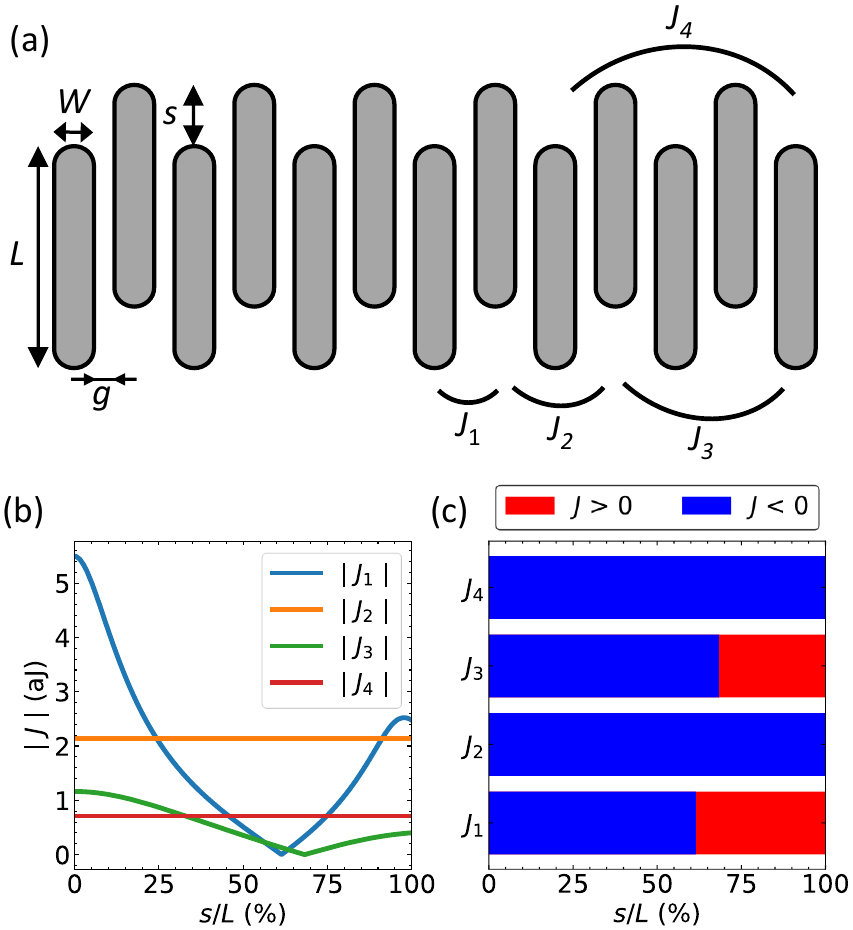}
\caption{(a) Illustration of a mesoscopic magnetic chain. The stadium-shaped elements have typical length $L$ = 1500 nm and width $W$~=~150~nm, while the gap between neighboring mesospins is $g$~=~75~nm. A periodic vertical shift $s$ is applied to every second element to tune the magnetic interactions. These interactions are described by the coupling strength $J_{i}$. (b) The amplitude and (c) sign of coupling strength for the first ($J_{1}$), second ($J_{2}$), third ($J_{3}$), and fourth ($J_{4}$) nearest neighbors calculated for Py material with MuMax$^{3}$ \cite{VansteenkisteArne2014Tdav}.}
\label{Figure1}
\end{figure}

The effect of the vertical shift $s$ on the dipolar interactions is explored performing micromagnetic simulations using MuMax$^{3}$ \cite{VansteenkisteArne2014Tdav}. The demagnetization energy was computed for a pair of mesospins coupled ferromagnetically $E_{\mathrm{fm}}$ (parallel alignment) and antiferromagnetically $E_{\mathrm{afm}}$ (antiparallel alignment), with a spacing between them corresponding to the pair of interest (first-, second-neighbors etc.) while the ratio $s/L$ varies from 0\% to 100\% with a step 0.2\%. Subsequently, we calculate the coupling strength as $J$ = $(E_{\mathrm{afm}}-E_{\mathrm{fm}})/2$ \cite{NguyenV.-D.2017Ciia}. 
To imitate the experimental conditions, the simulations were performed with stadium-shaped elements with dimensions 1500 $\times$ 150 $\times$ 20 nm$^{3}$. 
The gap between the mesospins is $g$ = 75 nm for the calculation of the coupling strength between the first nearest-neighbors $J_{1}$, and 2$g$ + $W$, 3$g$ + 2$W$ and 4$g$ + 3$W$ for the calculation of the coupling strength between the second $J_{2}$, third $J_{3}$ and fourth $J_{4}$ nearest-neighbors respectively.
For the simulations, we use Py as material.
Hence, the magnetic properties of mesospins, such as the saturation magnetization, $M_{sat}$ = 8.6 $\times$ 10$^{5}$ A/m, and the exchange stiffness constant, $A_{ex}$ = 13 $\times$ 10$^{-12}$ J/m, were chosen based on Py literature data \cite{VansteenkisteArne2014Tdav}. The mesospins were discretized into grid cells with cell size 2.5~nm. The magnetization within the elements is forced to be uniform, without allowing for any relaxation.


The magnetic state of the mesoscopic chains was investigated using real-space imaging. The Py conventional patterned chains were examined by performing Magnetic Force Microscopy (MFM), whereas the magnetic state of Fe$_{x}$Pd$_{100-x}$ implanted chains was determined by photoemission electron microscopy, employing x-ray magnetic circular dichroism (PEEM-XMCD) experiment at the CIRCE (bl24) beamline at the ALBA synchrotron \cite{AballeLucia2015TAsL}. Prior to microscopy, the chains were thermally demagnetized by heating them in ultra-high vacuum to a temperature above the Curie temperature of the ferromagnetic material, and subsequently cooled down with a controllable rate. The magnetic images were recorded far from the boundaries. 

We begin by discussing the effect of the geometrical vertical shift $s$ on a quasi-infinite chain. To this end, we examine the amplitude and sign of the coupling strength between the first ($J_{1}$), the second ($J_{2}$), the third ($J_{3}$), and the fourth ($J_{4}$) nearest-neighbors. The amplitude of coupling strength determines the strength of interactions, whereas its sign determines the ground state. Here we define that $J$ > 0 for ferromagnetic interaction (parallel alignment of neighboring mesospins) and $J$ < 0 for antiferromagnetic interaction (antiparallel alignment of neighboring mesospins). The calculations demonstrate that the vertical shift $s$ allows the tuning of the coupling strengths $J_{1}$ and $J_{3}$, in both amplitude (Fig. \ref{Figure1} (b)) and sign (Fig. \ref{Figure1} (c)).
As a consequence, the hierarchy of interactions varies, while conflicting coupling forces are present in the system. 

To explore the impact of the competing coupling forces on the magnetic ordering, we proceed by fabricating mesoscopic chains and performing real-space imaging. We fabricate conventional lithographic chains of Py material with six different vertical shifts ($s/L$ = 0\%, 8.7\%, 25\%, 50\%, 75\%, and 100\%), and image them with MFM after thermally demagnetizing them. Representative MFM images are displayed in Fig.~\ref{Figure2}~(a). The observed magnetic contrast confirms that the chains have mainly obtained two types of order:  antiferromagnetic (antiparallel alignment between neighboring mesosopins) for $s/L$~=~0\%, and dimer antiferromagnetic (antiparallel alignment between two neighboring pairs of mesosopins) for $s/L$ $\geq$ 25\%. 

The magnetic order of each chain is quantified by calculating the correlation number. This number is defined as: 
\begin{equation}
        C_{n} = \frac{\sum_{i} ^{N-n} m_{i} m_{i+n}}{N-n}
\end{equation}
where $n$ is the class of neighbors, i.e. $n$ = 1, 2, ... for the first, second, ... nearest-neighbors, $N$ is the total amount of elements, and $m_{i}$ and $m_{i + n}$ are the magnetization of the $i$-th and ($i+n$)-th element respectively ($m_{i}$ = 1 if the magnetization points upwards, or $m_{i}$ = -1 if the magnetization points downwards). The correlation number $C_{n}$ determines the extent to which $m_{i}$ and $m_{i + n}$ are related, and can take values from -1 to 1. $C_{n}$ = -1 ($C_{n}$ = 1) corresponds to
an antiparallel (parallel) alignment between $m_{i}$ and $m_{i + n}$, while $C_{n}$ = 0 indicates a random alignment. The correlation numbers for a full antiferromagnetic and dimer antiferromagnetic order are shown in Fig.\ref{Figure2} (b) and (c) respectively.  
The correlation numbers extracted from the MFM images are shown in Fig. \ref{Figure2} (d). Here, two trends are observed: one indicates antiferromagnetic order for $s/L$ = 0\%, whereas the other trend for $s/L$ $\geq$ 25\% resembles that of dimer antiferromagnetic order. The modified chain with $s/L$ = 8.7\% exhibits a negative correlation of the nearest mesospins, but no correlation for further neighbors.

\begin{figure}[t]
\includegraphics[width=1\linewidth]{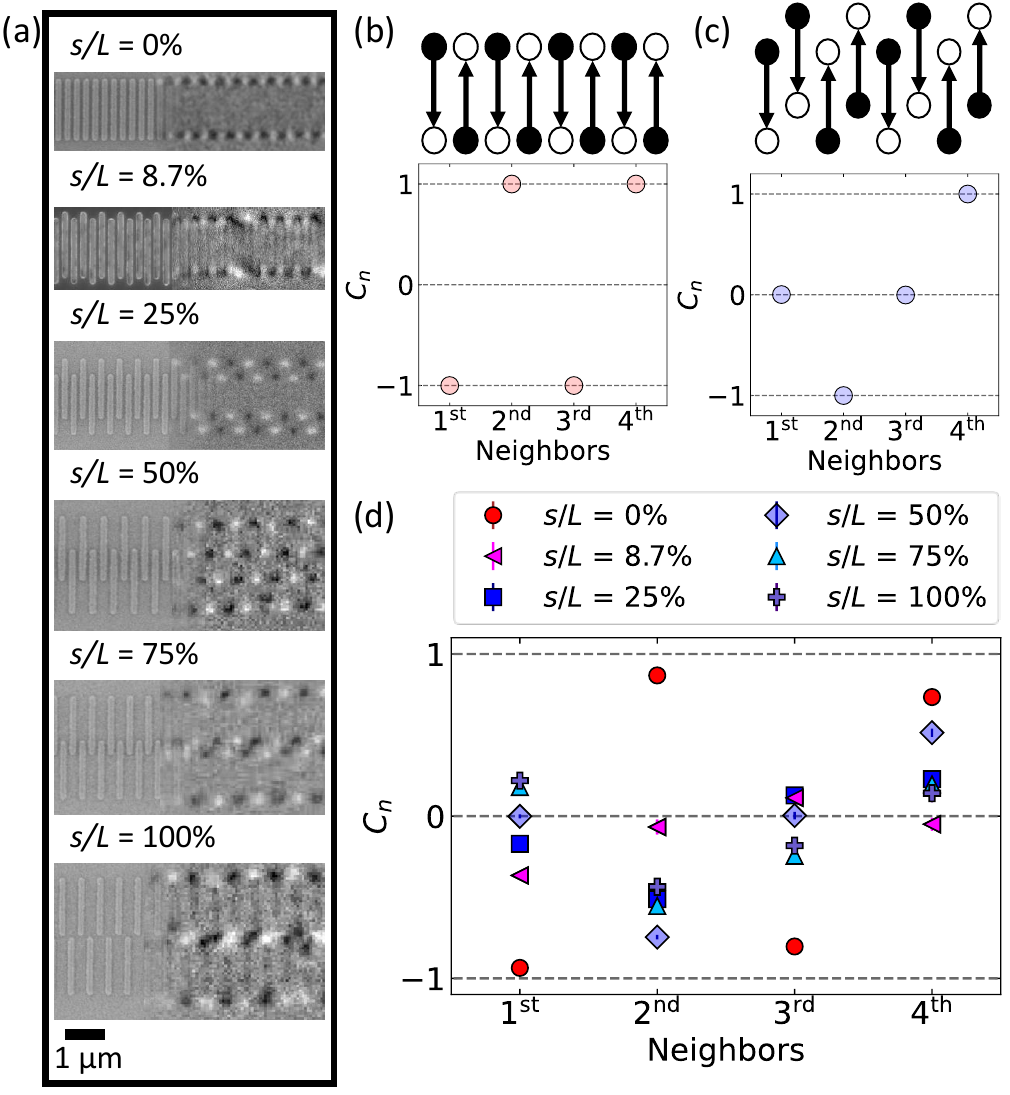}
\caption{(a) SEM (left) and MFM (right) images of the Py chains. The dark and bright regions correspond to the magnetic poles of the elements. (b,c) Correlation numbers for a full (b) antiferromagnetic and (c) dimer antiferromagnetic order. (d) Correlation numbers extracted from the MFM images of Py chains.}
\label{Figure2}
\end{figure}

Another way of studying the collective magnetic order is by calculating the Fast Fourier Transform (FFT) of the real-space images. For the FFT analysis, the number of elements per chain have been standardized to ensure uniform statistics. FFT intensities ($\left| \textrm{FFT} \right| ^{2}$) are presented as a function of the reciprocal lattice unit vector $\textbf{q}$ in Fig. \ref{Figure3}, where $q$ is the inverse of the magnetic periodicity. 
Two mirror-symmetric peaks are observed for $s/L$ = 0\% and $s/L \geq$ 25\%. The positions of the peaks reflect antiferromagnetic ordering for $s/L$~=~0 and dimer antiferromagnetic ordering for $s/L \geq$~25\%. Low-intensity peaks are also noticed at $q$ = 0 for $s/L$ = 75\% and 100\%. The $q$ = 0 peak indicates the formation of ferromagnetic domains, which reflect that the chains possess a net moment. The modified chain with $s/L$ = 8.7\% is characterized by a diffuse spectrum, and hence a disordered configuration in the real space. 

\begin{figure}[t]
\includegraphics[width=1\linewidth]{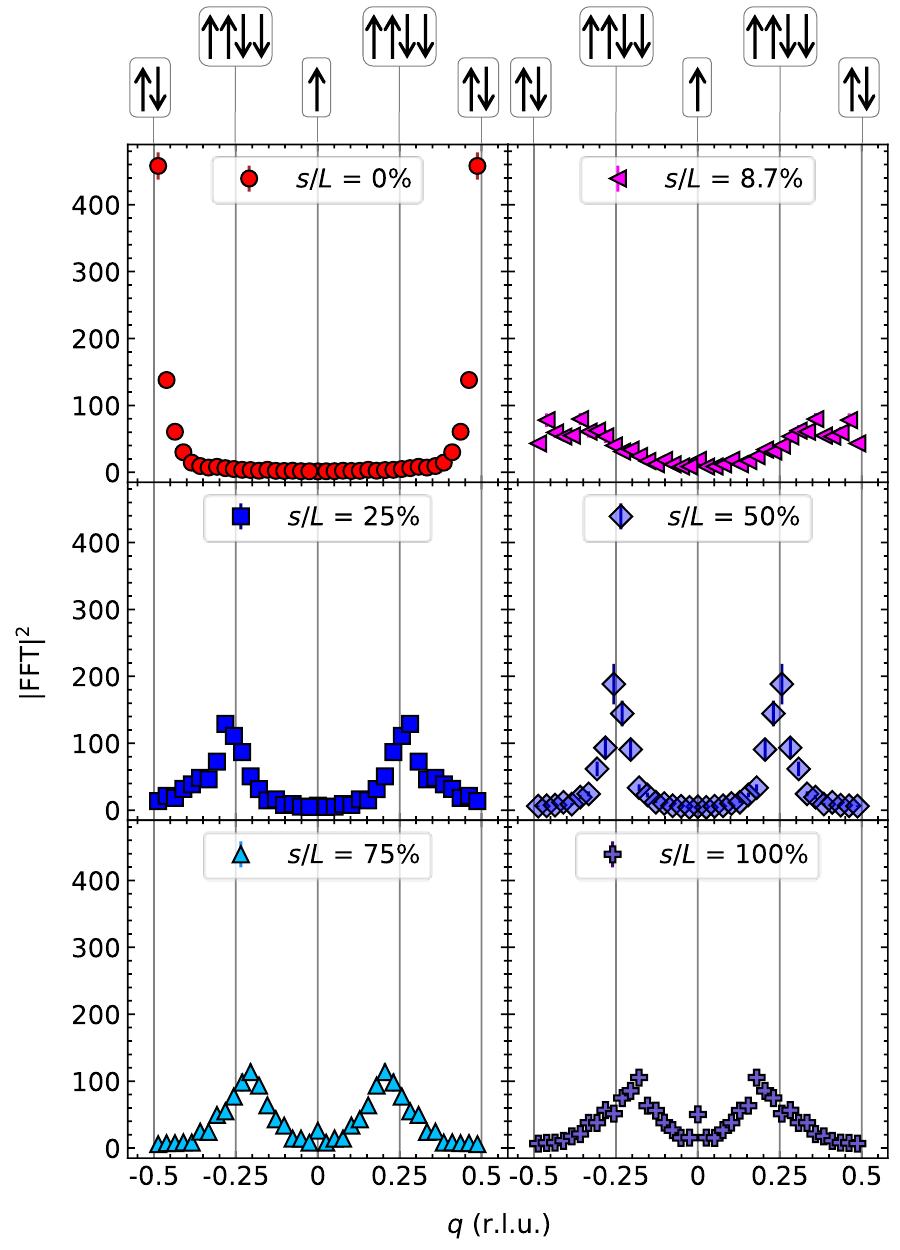}
\caption{FFT intensities as a function of the reciprocal lattice vector $q$ in reciprocal lattice units (r.l.u.) for the Py chains. The reciprocal lattice unit vector is defined by the structural periodicity $(W+g)$. The data show well-defined peaks for $s/L$ = 0\%, 25\%, 50\%, 75\% and 100\%, but a diffused spectrum for $s/L$ = 8.7\%, indicating ordered and disordered configurations respectively.}
\label{Figure3}
\end{figure}

To understand the role of interactions in the received ordering, we need to take a look at the total energy per mesospin. For the calculation of this energy, we make an analogy between mesospins and spins. This analogy allows us to consider that the total energy of a chain of $N$ mesospins is given by $E_{ij}$ = -$\sum_{ij} J_{ij} s_{i} s_{j}$, where here $s_{i}$ and $s_{j}$ are the magnetization of mesospins $i$ and $j$ that interact with energy $J_{ij}$. Consequently, taking into account the contribution of first to fourth-nearest neighbors (Fig.~\ref{Figure1}~(b) and (c)), the total energy for antiferromagnetic, dimer antiferromagnetic and ferromagnetic ordering is given by:  
\begin{equation}
E_{\mathrm{afm}} = 2(J_{1} - J_{2} + J_{3} - J_{4})
\label{Energy per spin afm}
\end{equation}
\begin{equation}
E_{\mathrm{dimer}} = 2(J_{2} - J_{4})
\label{Energy per spin dimer}
\end{equation}
\begin{equation}
E_{\mathrm{fm}} = -2(J_{1} + J_{2} + J_{3} + J_{4})
\label{Energy per spin fm}
\end{equation}
The results are summarized in Fig. \ref{Figure4}.
The antiferromagnetic ordering has the lowest energy for $s/L <$ 15.2\%, while for larger vertical shifts the dimer antiferromagnetic ordering is favoured. As evidenced by Eqs. \ref{Energy per spin afm}, \ref{Energy per spin dimer}, the emergent order is a consequence of the long-range interactions. The antiferromagnetic-dimer to antiferromagnetic transition point is close to the $s/L$ = 8.7\% case, providing insight into the diffuse FFT spectrum. 
It is worth noting that the phase diagram remains valid even if considering higher-order neighbors (see Supplemental Material).
To a good approximation, previous theoretical models have also predicted the observed phase diagram by considering competing interactions only between the nearest and next-nearest neighbors in spin chains \cite{SelkeWalter1988TAm}. 
It is also important to mention that the Eqs. \ref{Energy per spin afm}, \ref{Energy per spin dimer} and \ref{Energy per spin fm} confirm our experimental findings because the chains are quasi-infinite, and hence characterized by uniform interactions far from the boundaries. For finite-size chains, whose interactions are truncated, a multiphase diagram has been observed \cite{NguyenV.-D.2017Ciia}.


\begin{figure}[t]
\includegraphics[width=1\linewidth]{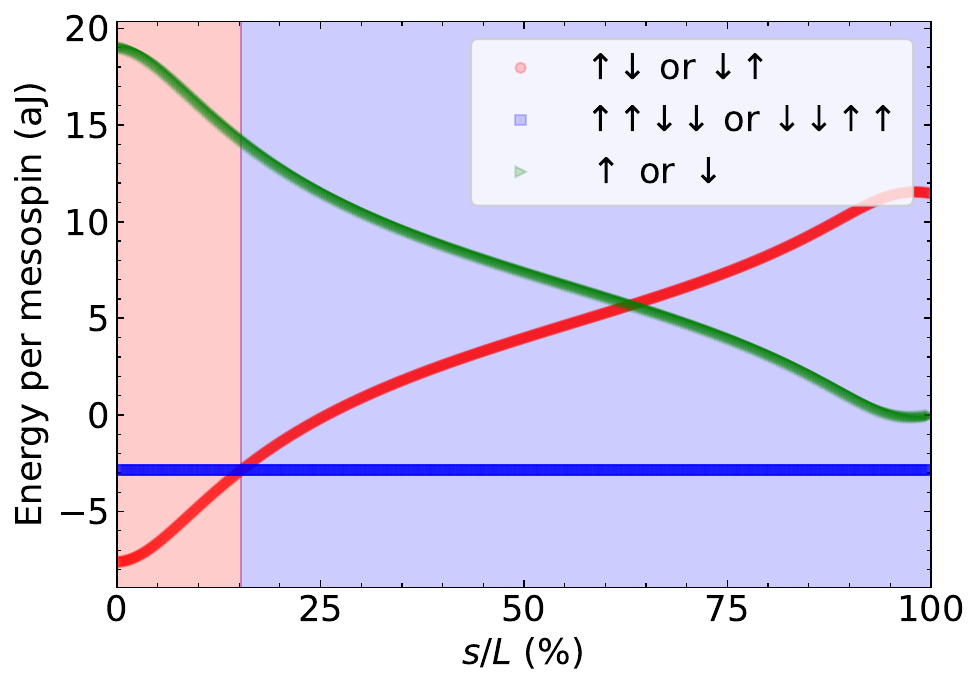}
\caption{Calculated total energy per mesospin at room temperature for the antiferromagnetic ($ \uparrow\downarrow$ or $\downarrow\uparrow$), dimer antiferromagnetic ($\uparrow\uparrow\downarrow\downarrow$ or $\downarrow\downarrow\uparrow\uparrow$) and ferromagnetic ($\uparrow$ or $\downarrow$) ordering. To calculate the energy per spin, we used the simulated coupling strengths for Py chains.}
\label{Figure4}
\end{figure}

To explore the extent of the impact of the geometrical vertical shift on the ordering, we choose to investigate the same chains made by a different material. In this system, the magnetic functionality of elements originates by the implantation of Fe$^{+}$ into a Pd film.
The implanted chains were imaged with XMCD-PEEM technique at room temperature after thermally demagnetizing them. 
The correlation numbers (Fig. \ref{Figure5}) and Fourier analysis (Fig. \ref{Figure5} and Supplemental Material) for the implanted chains are again in line with the antiferromagnetic - dimer to antiferromagnetic order phase diagram. However, in comparison to conventionally patterned chains, a different behavior is encountered for $s/L$ = 8.7\% and 25\%. Both real and reciprocal space analysis show antiferromagnetic ordering for $s/L$~=~8.7\%, and a less ordered state for $s/L$ = 25\%. This difference indicates the shift of the antiferromagnetic - dimer to antiferromagnetic transition point towards higher $s$ values. A possible reason for this shift is the inhomogeneous depth distribution of the implanted Fe within the Pd matrix resulting from the implantation process \cite{vantaraki2024magneticmetamaterialsionimplantation}. It is also important to mention that the implanted chains are characterized by lower correlation numbers and intensities in reciprocal space than the conventional patterned chains, indicating higher disorder. This difference might be a consequence of the lower magnetic moment of Fe$_{x}$Pd$_{100-x}$ compared to Py sample (for continuous films M$_{sat}^{\mathrm{Py}}$/M$_{sat}^{\mathrm{FePd}}$ $\sim$ 9), and hence weaker interactions.

\begin{figure}[t]
\includegraphics[width=1\linewidth]{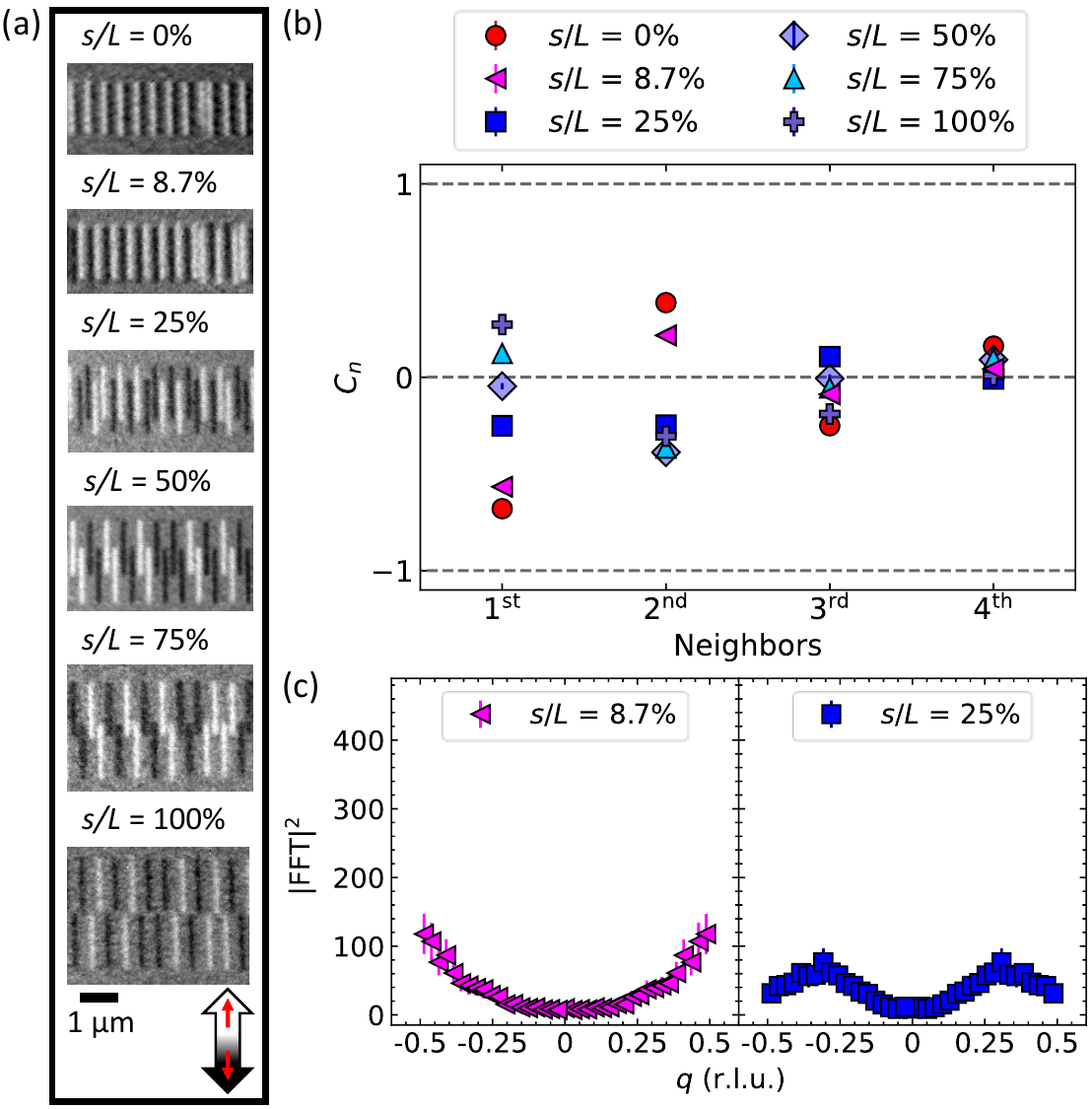}
\caption{(a) XMCD-PEEM images of the implanted chains. (b) Correlation numbers extracted from the XMCD-PEEM images for the implanted chains. (c) FFT intensities for the implanted chains with $s/L$ = 8.7\% and 25\%.}
\label{Figure5}
\end{figure}


While the observed magnetic configurations are static, they result from the thermal-driven dynamics during the annealing protocol. To investigate if the annealing protocols were effective for the chains to reach their ground state, we examine if the mesoscopic systems are in thermal equilibrium. To do so, we take a look at the energy distribution of states at the blocking temperature $T_{B}$, under which the mesospins become athermal (i.e. the magnetization does not fluctuate at the timescale of the measurements). For this calculation, we count the experimental probabilities from the real-space images, and compare them to the theoretical probabilities from the Boltzmann distribution at $T_{B}$. The experimental probabilities are measured at room temperature, but since the magnetic configuration is frozen, they are comparable to those at $T_{B}$. 
For the experimental probabilities, we count the populations of all possible five-elements magnetization permutations from the real-space images. These permutations include the 2$^5$ five-element magnetic configurations as $\uparrow\uparrow\uparrow\uparrow\uparrow$, $\uparrow\uparrow\uparrow\uparrow\downarrow$, $\uparrow\uparrow\uparrow\downarrow\downarrow$, ... . The energy of the $i$-th permutation is calculated as described above, and the corresponding ground-state energy for each $s$ is subtracted from it\textcolor{red}{,} yielding the relative permutation energy $E_{i}(T_R)$, where $T_R$ stands for room temperature. The populations are normalized by the total number of elements, yielding the experimental probabilities $\overline{p_{i}}$. 
If the system is at thermal equilibrium at $T_{B}$, the experimental probabilities will follow the Boltzmann distribution, which is given by:
\begin{equation}
p_i=\frac{1}{Z(T_{B})}\mathrm{exp}\left(\frac{-E_i(T_{B})}{k_B T_{B}}\right),
\end{equation}
where $Z(T_{B})= \frac{1}{k_B T_{B}}\sum_i E_i(T_{B})$ is the partition function at $T_{B}$, $E_i(T_{B})$ is the energy of the $i$-th permutation at $T_{B}$, and $k_{B}$ is the Boltzmann constant. Since only the energies at room temperature $E_{i}(T_R)$ are known, we consider $E_i(T_{B})$ to be proportional to $E_{i}(T_R)$ by a dimensionless factor $f_{E}$, so that $E_i(T_{B})$ = $f_{E} E_{i}(T_R)$ for each $s$. 
As the mesospins interact solely through dipolar interactions, these energies are proportional to $M_{sat}$. Thus, we expect that near the Curie temperature of the ferromagnetic material, the relative permutation energies $E_i(T)$ will decrease abruptly in a small temperature range, facilitating the magnetization reversal of the mesospins.
Therefore, we can assume $T_{B}$ to be equal to the Curie temperature of the ferromagnetic material ($T_{C}^{Py}$ = 790 K \cite{ZhangXiaoyu2019Utao} and $T_{C}^{FePd}$ = 335 K) and we extract the $f_{E}$ for each $s$ from fitting $p_i$ to $\overline{p_{i}}$ (see Fig. \ref{Figure6} and Supplemental Material for the individual fits). The fitting yields an average factor $\overline{f_{E}}$=0.005(1) for the conventional patterned chains and $\overline{f_{E}}$=0.0011(1) for the implanted chains.
As $f_{E} \ll$ 1 for each $s$, we conclude that the energy differences at $T_B$ are considerably smaller compared to the ones at room temperature. This result indicates a $T_B$ close to $T_{C}$, confirming our original hypothesis and showing that the hierarchy of interactions holds up even close to $T_C$. In addition, the good agreement of the data with the exponential decay (see Fig. \ref{Figure6}) shows that the mesoscopic chains are in thermal equilibrium at $T_B$.

\begin{figure}[t]
\includegraphics[width=1\linewidth]{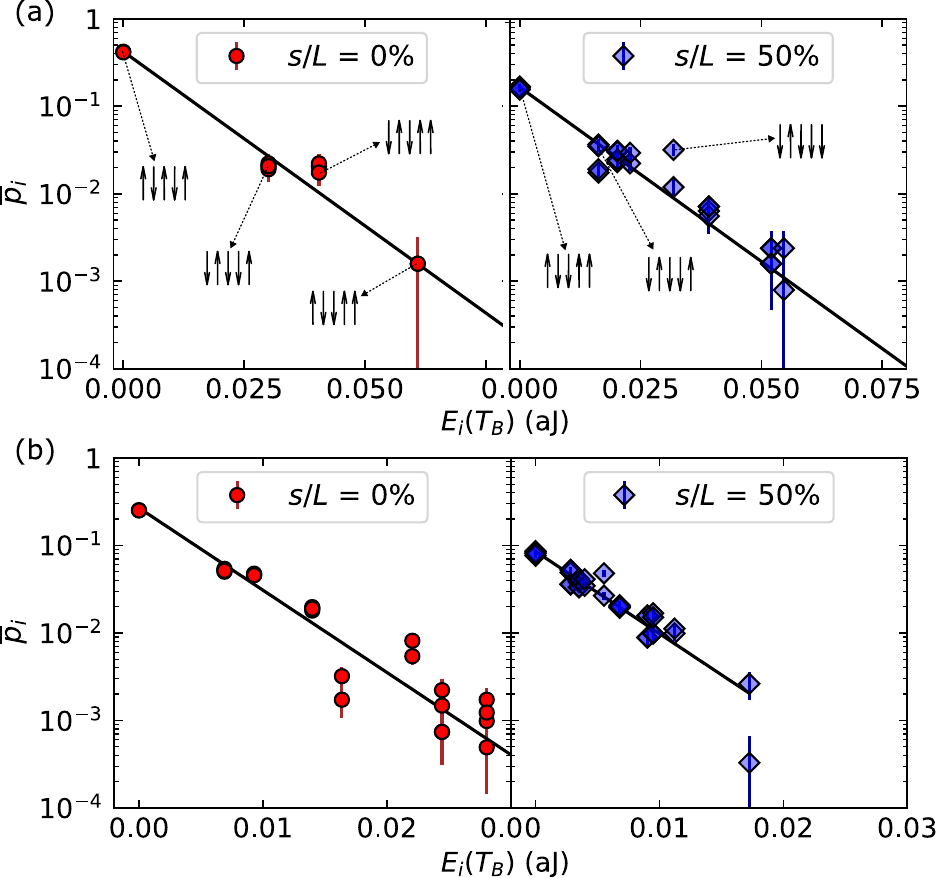}
\caption{Experimental probabilities of the permutation $\overline{p_{i}}$ as a function of the energies at the blocking temperature $E_{i}(T_B) = f_{E} E_{i}(T_R)$ for (a) conventional patterned and (b) implanted chains with $s/L=$ 0\% and 50\%. The solid lines show the fitted Boltzmann distributions $p_i$.}
\label{Figure6}
\end{figure}

In conclusion, we have demonstrated that mesoscopic magnetic chains exhibit two distinct ground states, depending on the shift parameter $s$ modifying the interactions: an antiferromagnetic phase and a dimer antiferromagnetic phase. These two configurations arise from the competition between long-range magnetic interactions and an energy hierarchy that remains robust even near the Curie point. While the focus of this study is on fundamental physics, our results reveal that this mesoscopic system autonomously converges to a low-energy state from an extended landscape of metastable configurations, in a fashion similar to addressing a combinatorial optimization challenge \cite{Bybee2023, MohseniNaeimeh2022Imah}. In essence, these magnetic chains behave analogously to a physical Ising machine, providing a tangible platform for solving complex optimization problems \cite{Wang2021_Ising, SiJia2024EsIm}. Looking ahead, future research could build on these findings by exploring the influence of long-range interactions on chains of varying sizes, which would not only deepen our understanding of long-range magnetic interactions but also define the operational limits of such Ising machine analogs. Investigating these systems under different geometric configurations or environmental conditions could further illuminate the boundaries between fundamental magnetic phases and their applicability in solving real-world optimization problems. This will potentially pave the way for practical implementations of mesoscopic Ising machines in computational tasks, as demonstrated previously by \citet{Bhanja_NatNano_2016}.

\begin{acknowledgments}
The authors would like to thank Johan Oscarsson and Mauricio Sortica at the Uppsala Tandem Laboratory for help with ion implantations. The authors are thankful for an infrastructure grant by VR-RFI (grant number 2019-00191) supporting the accelerator operation. The authors also acknowledge support from the Swedish Research Council (projects no. 2019-03581 and 2023-06359). CV gratefully acknowledges financial support from the Colonias-Jansson Foundation, Thelin-Gertrud Foundation, Liljewalch and Sederholm Foundation. CV and VK would like to
express their gratitude to Dr. Sam D. Slöetjes for fruitful discussions. KI and UBA acknowledge funding from the Icelandic Research Fund project 2410333. We acknowledge Myfab Uppsala for providing facilities and experimental support. Myfab is funded by the Swedish Research Council (2020-00207) as a national research infrastructure. 
The PEEM-XMCD experiments were performed at CIRCE (bl24) beamline at ALBA Synchrotron with the collaboration of ALBA staff. M.F. acknowledges support from MICIN through grant number PID2021-122980OB-C54.
The authors are deeply grateful for the support from ReMade@ARI, funded by the European Union as part of the Horizon Europe call HORIZON-INFRA-2021-SERV-01 under grant agreement number 101058414 and co-funded by UK Research and Innovation (UKRI) under the UK government’s Horizon Europe funding guarantee (grant number 10039728) and by the Swiss State Secretariat for Education, Research and Innovation (SERI) under contract number 22.00187.

\end{acknowledgments}

\hfill \break
\noindent

The data that support the findings of this study are available from the corresponding authors upon reasonable request.


%

\pagebreak
\onecolumngrid
\newpage
\begin{center}
\textbf{\large Supplemental Material: Magnetic metamaterials by ion-implantation}
\end{center}
\setcounter{equation}{0}
\setcounter{figure}{0}
\setcounter{table}{0}
\setcounter{page}{1}
\makeatletter
\renewcommand{\theequation}{S\arabic{equation}}
\renewcommand{\figurename}{Supplementary FIG.}
\renewcommand{\thefigure}{{\bf \arabic{figure}}}
\renewcommand{\bibnumfmt}[1]{[S#1]}
\renewcommand{\citenumfont}[1]{S#1}
\renewcommand{\thepage}{S-\arabic{page}}


\title{Supplemental Material: Magnetic Order and Long-Range Interactions in Mesoscopic Ising Chains}

\maketitle
\clearpage

\section{Energy per mesospin}
To assess the robustness of the results presented in Fig. 4 when considering interactions beyond the fourth neighbor, we analytically calculate the energy per mesospin for infinite chains as a function of the shift $s$. For this analysis, we assume that the coupling strengths decay with distance according to a power-law function, $y(x) = ax^b$. Using this relationship, we fit the coupling strengths $J$ for odd and even $n$ neighbors at each value of $s$ (see Supplementary Fig. \ref{AppendixA} (a) and (b) for the case of $s/L = 0$), as follows: 
\begin{equation}
J_{n} = an^{b} \textrm{}\ \textrm{for}\ n = 1, 3, 5, ...
\label{Jodd}
\end{equation}
\begin{equation}
J_{n} = a^{\prime} n^{b^{\prime}} \textrm{}\ \textrm{for}\ n = 2, 4, 6, ...  
\label{Jeven}
\end{equation}
Consistent with Eqs. (2), (3), and (4), the total energy per mesospin for antiferromagnetic, dimer antiferromagnetic, and ferromagnetic orderings is given by:
\begin{equation}
E_{\mathrm{afm}} = 2 \left[ a \left(\sum_{n=1,3,5,\ldots}^{\infty} n^{b}\right) - a' \left(\sum_{n=2,4,6,\ldots}^{\infty} n^{b'}\right) \right]
\label{Energy afm}
\end{equation}
\begin{equation}
E_{\mathrm{dimer}} = 2 a' \left[\sum_{\substack{n=2,4,6,..., \\ n \textrm{}\ \mathrm{mod} \textrm{}\ 4 \neq 0 }}^{\infty} n^{b'} - \sum_{\substack{n=2,4,6,..., \\ n \textrm{}\ \mathrm{mod} \textrm{}\ 4 = 0 }}^{\infty} n^{b'}\right]
\label{Energy dimer}
\end{equation}
\begin{equation}
E_{\mathrm{fm}} = -2 \left[a\left(\sum_{n=1,3,5,...}^{\infty} n^{b}\right) + a' \left(\sum_{n=2,4,6,...}^{\infty} n^{b'}\right)\right]
\label{Energy fm}
\end{equation}
Numerical results are presented in Supplementary Fig. \ref{AppendixA} (c) for $n = 10$, 20, and 30. For small values of $s/L$, the antiferromagnetic state remains energetically favorable, while at larger $s/L$, a transition to the dimer antiferromagnetic state is observed, consistent with the findings in Fig. 4.
It is important to note that the power-law scaling does not hold for intermediate shifts ($s/L > 50$\%), where the hierarchy of interaction energies changes (e.g., $|J_{3}| \geq |J_{1}|$). As a result, Eqs. \ref{Energy afm}, \ref{Energy dimer}, and \ref{Energy fm} are not applicable in this range of $s/L$.
\begin{figure}[!h]
\includegraphics[width=1\linewidth]{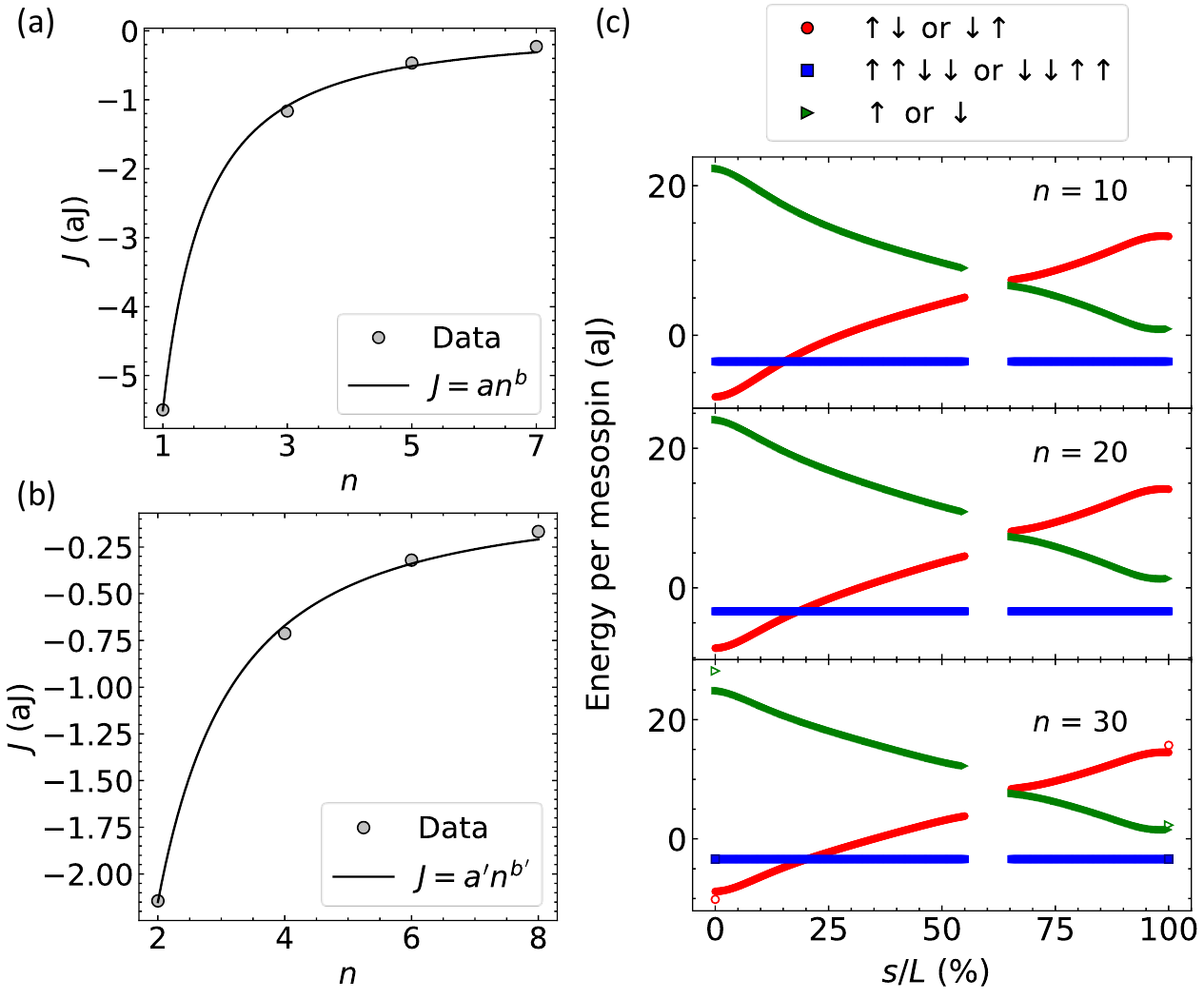}
\caption{The coupling strengths for (a) odd and (b) even $n$ neighbors for the chain with $s/L$ = 0\%. The data represent micromagnetic estimates of $J$, calculated as discussed in the main text. (c) The energy per mesospin for antiferromagnetic, dimer antiferromagnetic and ferromagnetic ordering for $n$ = 10, 20 and 30. The open markers in the $n$~=~30 plot represent the energies for an infinite number of neighbors.}
\label{AppendixA}
\end{figure}

\section{FFT intensities}
The FFT intensities for the implanted chains with $s/L$ = 0\%, 50\%, 75\%, and 100\% are presented in Supplementary Fig. \ref{Appendix1}. Similar to conventional patterned chains, the position of the peaks indicate antiferromagnetic ordering for $s/L$ = 0\% and dimer antiferromagnetic ordering for the rest.

\begin{figure}[!h]
\includegraphics[width=0.9\linewidth]{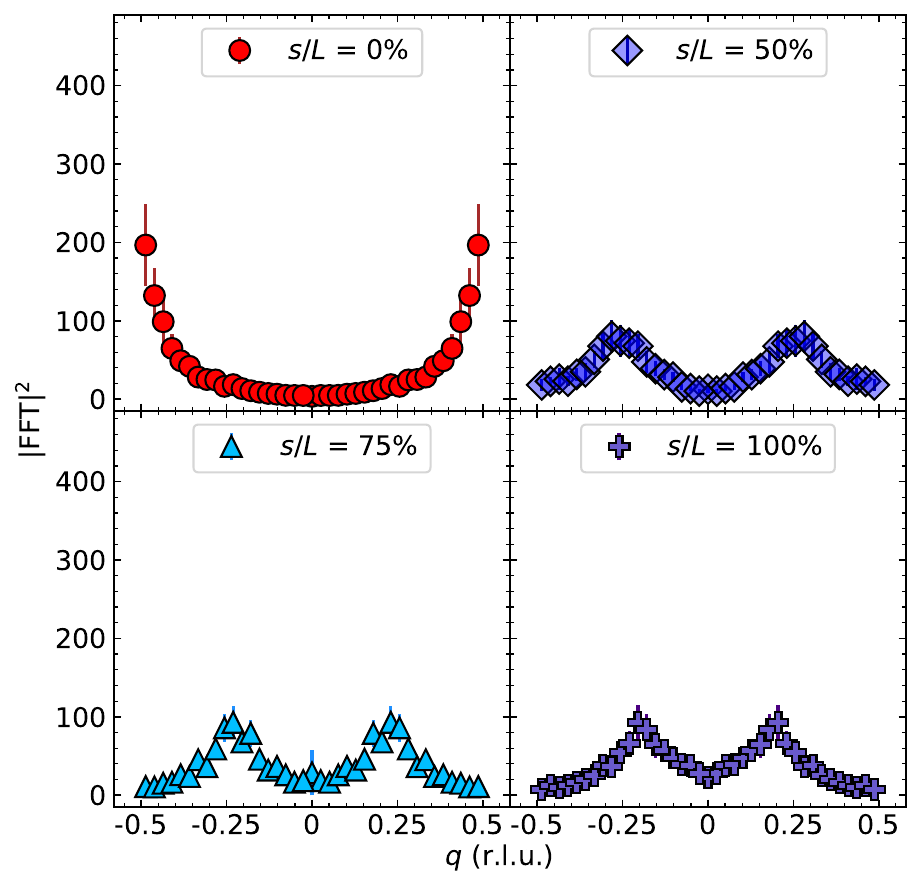}
\caption{FFT intensities for the implanted chains with $s/L$ = 0\%, 50\%, 75\%, and 100\%.}
\label{Appendix1}
\end{figure}

\clearpage
\section{Boltzmann distribution}
The experimental probabilities $\overline{p_{i}}$ and the fitted Boltzmann distribution for the conventional patterned and the implanted chains are shown in Supplementary Fig. \ref{Appendix_both} (a) and (b) respectively. For all chains, we consider that the blocking temperature is very close to and practically equal to the respective Curie temperature, and subsequently we fit a correction parameter $f_E$ (for each shift $s$). It is important to mention that the permutation energy depends on the shift $s$. For example, the ground states for $s/L=0$\% (where $E_i(T_B)=0$) correspond to $\uparrow\downarrow\uparrow\downarrow\uparrow$ and
$\downarrow\uparrow\downarrow\uparrow\downarrow$, while the ground states for $s/L=100$\% correspond to configurations with the symmetry of $\uparrow\downarrow\downarrow\uparrow\uparrow$.
The experimental probabilities of symmetric configurations in respect to the dipolar interactions could be superimposed in Supplementary Fig. \ref{Appendix_both}. For example, for $s/L=0$\%, the two antiferromagnetic populations are almost identical and their points are superimposed.
It is also worthy to note that the ground-state populations depend mostly on the energy gap to the next permutation. For example, the $s/L=0$\% case has the largest gap between the lowest-energy permutations, leading to a larger experimental probability $p_0=0.4$ compared with the other shifts. 
The error bars of $\overline{p_{i}}$ represent one standard error (SE) for a multinomial distribution, i.e., SE = $\sqrt{\overline{p_{i}}(1-\overline{p_{i}})/N}$.

\begin{figure}[!h]
\includegraphics[width=0.5\linewidth]{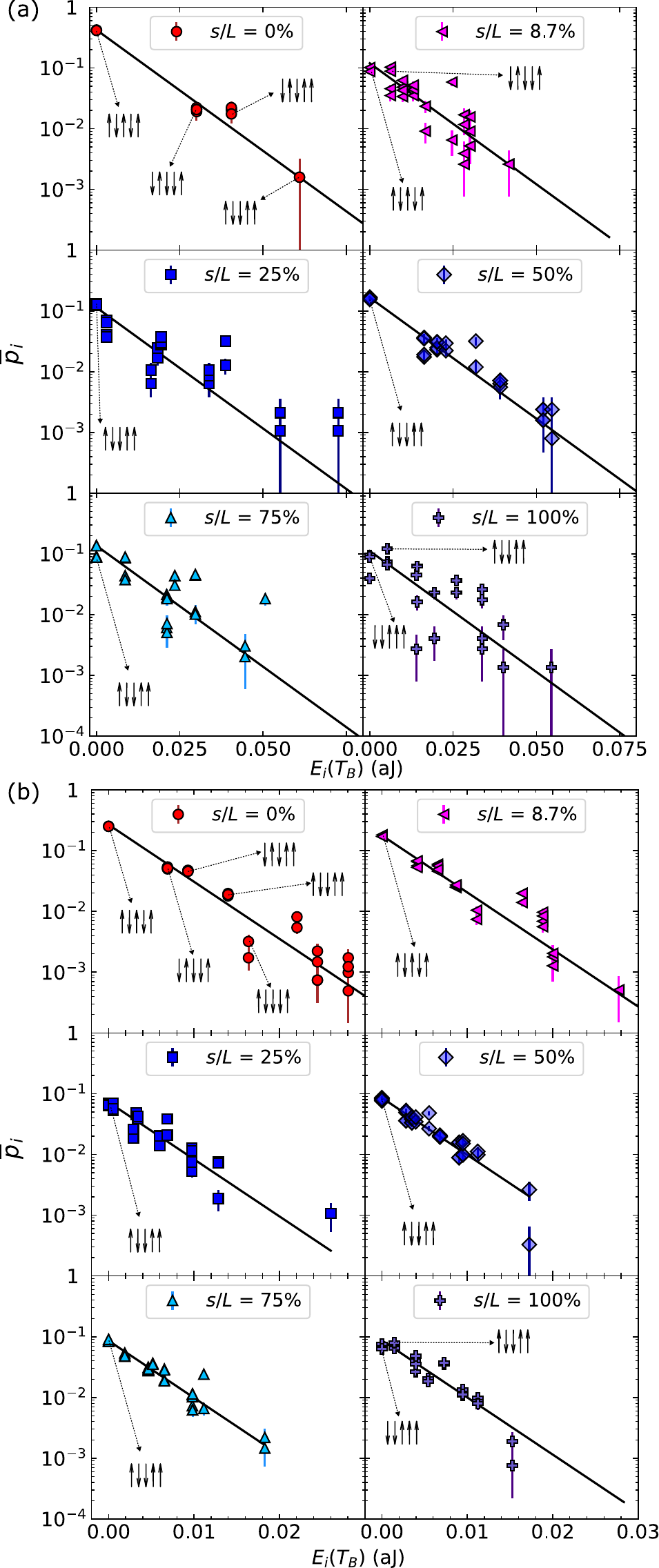}
\caption{Experimental probabilities $\overline{p_{i}}$ as a function of the energies at the blocking temperature $E_{i}(T_B) = f_{E} E_{i}(T_R)$ for (a) conventional patterned and (b) implanted chains. The continuous lines show the fitted Boltzmann distributions. 
The arrows in the plot denote the magnetic configurations corresponding to particular data points.}
\label{Appendix_both}
\end{figure}

\end{document}